\pdfoutput=1
\documentclass[a4paper]{article}

\usepackage[ninepoint]{itgspeech2021}    
\usepackage{times}            
\usepackage[english]{babel}   
\usepackage[utf8]{inputenc}
\usepackage[T1]{fontenc}      
\usepackage[sort&compress,numbers]{natbib}	
\usepackage{amsmath,amssymb}
\usepackage{graphicx}
\usepackage[colorlinks=false,pdfborder={0 0 0}]{hyperref}
\usepackage{units}
\usepackage{microtype}

\DeclareMathAlphabet{\mathcal}{OMS}{cmsy}{m}{n}

\sloppypar

\usepackage{amsmath}
\usepackage{amssymb}
\usepackage{bbm}
\usepackage{physics}
\usepackage{siunitx}
\usepackage{trfsigns}
\usepackage{pifont}
\usepackage{etoolbox}
\usepackage{multirow}
\usepackage{csquotes}
\usepackage[shortlabels]{enumitem}  
\usepackage{hyperref}
\usepackage[capitalize]{cleveref}
\usepackage{tikz}
\usepackage{balance} 
\usepackage{booktabs}

\usepackage{url}

\newcommand{\matr}[1]{\ensuremath{\boldsymbol{\mathbf{#1}}}}
\newcommand{\vect}[1]{\ensuremath{\boldsymbol{\mathbf{#1}}}}
\def\L{\ensuremath{\mathcal{L}}}
\def\J{\ensuremath{\mathcal{J}}}
\def\bigo{\mathcal{O}}
\newcommand*{\tran}{\smash{{}^{\mkern-1.5mu\mathsf{T}}}}
\newcommand*{\rawtran}{^{\mkern-1.5mu\mathsf{T}}}

\newcolumntype{H}{>{\setbox0=\hbox\bgroup}c<{\egroup}@{}}   

\newcommand*{\thl}{\fontseries{b}\selectfont}
\robustify\thl
\robustify\fontseries

\newcommand{\loss}[1]{\mathcal{L}^{\text{(#1)}}}
\newcommand{\jloss}[1]{\mathcal{J}^{\text{(#1)}}}

\usepackage{pgfplots}
\pgfplotsset{compat=1.16}
\usepgfplotslibrary{groupplots}
\usetikzlibrary{positioning,arrows,matrix,fit,calc,patterns,chains,scopes,shapes.multipart,decorations,decorations.markings,backgrounds}

\definecolor{palette-1}{rgb}{0.282352941176471,0.470588235294118,0.815686274509804}
\definecolor{palette-2}{rgb}{0.933333333333333,0.52156862745098,0.290196078431373}
\definecolor{palette-3}{rgb}{0.415686274509804,0.8,0.392156862745098}
\definecolor{palette-4}{rgb}{0.83921568627451,0.372549019607843,0.372549019607843}
\definecolor{palette-5}{rgb}{0.584313725490196,0.423529411764706,0.705882352941177}
\definecolor{palette-6}{rgb}{0.549019607843137,0.380392156862745,0.235294117647059}
\definecolor{palette-7}{rgb}{0.862745098039216,0.494117647058824,0.752941176470588}
\definecolor{palette-8}{rgb}{0.835294117647059,0.733333333333333,0.403921568627451}
\definecolor{palette-9}{rgb}{0.509803921568627,0.776470588235294,0.886274509803922}

\tikzset{
    line/.style={draw,black,thick,rounded corners=1mm,line cap=round},
    noshortarrow/.style={line,->},
    arrow/.style={noshortarrow,shorten >=.3mm},
    doublearrow/.style={arrow,<->, shorten <=.3mm},
    box/.style={draw,black,thick,minimum height=3em,text height=1.5ex,text depth=0.25ex,rounded corners=3,fill=white},
    nopadding/.style={minimum height=0,inner sep=1mm},
    signalbox/.style={draw,black,thin,rounded corners=1mm,minimum width=7mm, minimum height=4mm,inner sep=0},
    pbox/.style={box,fill=black!10},
    backgroundbox/.style={inner xsep=3mm, inner ysep=1mm, draw, dashed, rounded corners,fill=orange!10},
    branch/.style={inner sep=0.3mm,circle,fill=black},
    operator/.style={draw,circle,black,rounded corners,inner sep=0,fill=white},
    vertex/.style={draw,ultra thin,circle,black,rounded corners,inner sep=0.6mm,fill=gray,fill opacity=0.5},
    edge/.style={line,very thick,line cap=butt},
    pattern1/.style={pattern=north west lines,pattern color=palette-1},
    pattern2/.style={pattern=north east lines,pattern color=palette-2},
    pattern3/.style={pattern=crosshatch,pattern color=palette-3},
    buswidth/.style={path picture={\draw[black,-] (path picture bounding box.south west) -- (path picture bounding box.north east);}}
}

\makeatletter
\renewcommand\section{\@startsection {section}{1}{\z@}%
  {-1.5ex \@plus -0.6ex \@minus -0.1ex}{6pt \@plus.1ex}{\normalfont\Large\bfseries}}
\renewcommand\subsection{\@startsection{subsection}{2}{\z@}%
  {-1.4ex\@plus -0.5ex \@minus -0.1ex}{4pt \@plus 0.1ex}{\normalfont\large\bfseries}}
\renewcommand\subsubsection{\@startsection{subsubsection}{3}{\z@}%
  {-1.05ex\@plus -0.5ex \@minus -0.1ex}{2pt \@plus 0.1ex}{\normalfont\normalsize\bfseries}}
\makeatother

\usepackage[acronym,shortcuts]{glossaries}
\glsdisablehyper
\newacronym{SDR}{SDR}{Signal-to-Distortion Ratio}
\newacronym{CSS}{CSS}{Continuous Speech Separation}
\newacronym{PIT}{PIT}{Permutation Invariant Training}
\newacronym{uPIT}{uPIT}{Utterance-level \gls{PIT}}
\newacronym{MSE}{MSE}{Mean Squared Error}
\newacronym{DFS}{DFS}{Depth First Search}
\newacronym{DPRNN}{DPRNN}{Dual-Path Recurrent Neural Network}
\newacronym{WER}{WER}{Word Error Rate}
\newacronym{SSE}{SSE}{Sum Squared Error}
\newacronym{SI-SDR}{SI-SDR}{Scale-Invariant \gls{SDR}}
\newacronym{DP}{DP}{Dynamic Programming}

\title{Speeding Up Permutation Invariant Training for Source Separation}

\def\upb{$^1$}
\def\ntt{$^2$}
\author{Thilo von Neumann,\upb{} Christoph Boeddeker,\upb{} Keisuke Kinoshita,\ntt{} Marc Delcroix,\ntt{} \\ Reinhold Haeb-Umbach\upb}

\address{\upb Paderborn University, Germany; \ntt NTT Corporation, Japan\\
  Email: \texttt{\{vonneumann,boeddeker,haeb\}@nt.upb.de %
  }}

\begin{document}

\maketitle

\begin{abstract}
Permutation invariant training (PIT) is a widely used training criterion for neural network-based source separation, used for both utterance-level separation with utterance-level PIT (uPIT) and separation of long recordings with the recently proposed Graph-PIT.
When implemented naively, both suffer from an exponential complexity in the number of utterances to separate, rendering them unusable for large numbers of speakers or long realistic recordings.
We present a decomposition of the PIT criterion into the computation of a matrix and a strictly monotonously increasing function so that the permutation or assignment problem can be solved efficiently with several search algorithms.
The Hungarian algorithm can be used for uPIT and we introduce various algorithms for the Graph-PIT assignment problem to reduce the complexity to be polynomial in the number of utterances.
\end{abstract}



\def\aSDR{a-SDR}
\def\saSDR{sa-SDR}
\def\numutt{U}
\def\uttidx{u}
\def\numspk{K}
\def\spkidx{k}
\def\numchn{C}
\def\chnidx{c}
\def\time{T}
\def\timeidx{t}
\def\target{\vect{s}}
\def\estimate{\hat{\vect{s}}}
\def\meeting{\vect{y}}
\def\mixture{\vect{y}}
\def\spkuttset{\mathcal{U}^\text{(spk)}}
\def\chnuttset{\mathcal{U}^\text{(chn)}}
\def\assignment{\pi}
\def\permset{\mathcal{P}}
\def\permsetf{\Pi_\numchn}
\def\perm{\pi}
\def\permmatrix{\matr{P}}
\def\upittarget{\target^{\text{(uPIT)}}}
\def\targetmatrix{\matr{S}}
\def\estimatematrix{\hat{\matr{S}}}
\def\outerfn{f}
\def\lossmatrix{\matr{M}}
\def\lossmatrixentry{m}
\def\graph{\mathcal{G}}
\def\vertexset{\mathcal{V}}
\def\edgeset{\mathcal{E}}
\def\gpittarget{\target^\text{(Graph-PIT)}}
\def\coloringset{\mathcal{B}_{\graph,\numchn}}
\def\coloringmatrix{\permmatrix}
\def\adjacencymatrix{\matr{A}_\graph}

\section{Introduction}

Speech source separation is an important pre-processing step when applying speech recognition or diarization to realistic recordings of meeting scenarios.
It aims at estimating speech signals of individual speakers from a speech mixture. 
We can in general discern utterance-level algorithms \cite{Kolbaek2017_MultitalkerSpeechSeparationa,Luo2018_TaSNetTimeDomainAudio,Luo2020_DualPathRNNEfficient} and \gls{CSS} \cite{Chen2020_ContinuousSpeechSeparation,Yoshioka2018_RecognizingOverlappedSpeech} algorithms.
Utterance-level separation works on relatively short recordings of the length of roughly one utterance and aims at outputting each source on an individual output channel.
\gls{CSS} algorithms operate on a continuous stream of audio of arbitrary length and allow output channels to contain different speakers as long as they do not overlap.

Recently, neural networks have shown strong separation performance,  even if only single-channel recordings are available \cite{Luo2018_TaSNetTimeDomainAudio,Luo2019_ConvTasNetSurpassingIdeal,Luo2020_DualPathRNNEfficient}. 
A state-of-the-art technique is \gls{PIT} of a separation network.
\gls{uPIT} \cite{Kolbaek2017_MultitalkerSpeechSeparationa} finds the optimal permutation between the target signals and the network output channels during training.
This gives the network the freedom of selecting an output channel for each speaker.

We can use a network trained with uPIT to also realize CSS by use of a stitching scheme \cite{Yoshioka2018_RecognizingOverlappedSpeech,Chen2020_ContinuousSpeechSeparation} during test time.
The input audio stream is segmented into relatively short, temporally overlapping segments so that separation can be performed on each segment independently.
Neighboring segments are aligned using a similarity metric, which is computed on regions that are common to neighboring segments.
The maximum number of active speakers must not exceed the number of output channels in any segment.
This effectively limits the maximum segment size, because the number of output channels is typically chosen to be small, e.g., equal to two, and in multi-party meetings, long segments would typically comprise more than two speakers.
The alignment of adjacent segments also introduces additional computational overhead.

Recently, a technique called Graph-PIT \cite{vonNeumann2021_GraphPITGeneralizedPermutation} was introduced to relax this limitation by training a neural network to output more than one speaker on a single output channel.
Thus, Graph-PIT gives the freedom to increase the segment size of stitching-based \gls{CSS} significantly without loss in performance, reducing the computational overhead, or even eliminating stitching completely in certain scenarios.
We here focus only on these scenarios, i.e., we do not consider stitching.
Graph-PIT works by placing separated utterances on output channels solely under the constraint that 
overlapping utterances are placed on different output channels.
This transforms the permutation problem of \gls{uPIT} to a more general many-to-one assignment of utterances to output channels which we formulate as a graph coloring problem, as visualized in \cref{fig:gpit}.

However, graph coloring is an NP-hard problem and thus has an exponential complexity \cite{Bollobas1979_Colouring}.
While the complexity is not problematic for small models with two output channels, it explodes for more outputs, limiting the use of the original formulation \cite{vonNeumann2021_GraphPITGeneralizedPermutation} to small numbers of output channels and short recordings.

Both, \gls{uPIT} and Graph-PIT, have a factorial or exponential complexity in the number of utterances in the mixture signal when implemented naively \cite{Kolbaek2017_MultitalkerSpeechSeparationa,vonNeumann2021_GraphPITGeneralizedPermutation}.
It was shown that for \gls{uPIT}, the optimal assignment of target signals to output channels can be found with the Hungarian algorithm \cite{Kuhn_HungarianMethodAssignment,Munkres1957_AlgorithmsAssignmentTransportation} in polynomial time.
This makes it practical to train a model to separate up to 20 speakers \cite{Dovrat2021_ManySpeakersSingleChannel}.
A similar approach is not possible for Graph-PIT.

We show that an elegant decomposition of the loss function into a score matrix and a strictly monotonously increasing function always allows us to use more efficient algorithms for solving the \gls{uPIT} and the Graph-PIT assignment problems.\footnote{Code is available at \url{https://github.com/fgnt/graph_pit}}
We review how the Hungarian algorithm can be used to solve the permutation problem in \gls{uPIT} in polynomial time, and present a variety of assignment algorithms for Graph-PIT.
We found that the assignment problem can be formulated in a way such that dynamic programming can be applied to provide a solution in linear time in the number of utterances.
\section{Problem Formulation}

We consider the task of source separation, i.e., separating a speech mixture signal $\mixture$ containing overlapping speech of $\numspk$ speakers into its individual speech components.
We here only consider single-channel separation for simplicity, but the presented algorithms can as well be used to train multi-channel systems.
The speakers utter $\numutt$ utterances $ \targetmatrix = [\target_1, \target_2, ..., \target_\numutt] \in \mathbb{R}^{\time\times\numutt}$ where each utterance signal $\target_\uttidx\in\mathbb{R}^\time$ is represented as a vector of samples and zero-padded to the full length of the meeting $\time$.
The meeting signal $\meeting$ is the sum of these zero-padded utterance signals
\begin{align}
    \meeting = \sum_{\uttidx=1}^\numutt \target_\uttidx.
\end{align}

The general task is to separate $\meeting$ into $\numchn$ estimated signals $\estimatematrix = [\estimate_1, \estimate_2, ..., \estimate_\numchn] \in \mathbb{R}^{\time\times\numchn}$ so that no speech parts overlap.
We here only consider neural network-based separation with \gls{PIT} objectives where the network has a fixed number of $\numchn$ output channels and is trained with a \gls{SDR}-based loss function.
We look into two scenarios: utterance-level and meeting-level separation.

\paragraph{Utterance-level separation}
Utterance-level separation targets mixtures in which each speaker says exactly one utterance, and a separation system with $\numchn=\numspk$ output channels.
The task here is to place one utterance $\target_\spkidx$ onto each output channel.

\paragraph{Meeting-level separation}
The case of meeting-level \gls{CSS} allows for $\numspk > \numchn$ as long as never more than $\numchn$ speakers speak at the same time.
Exclusive placement of utterance different output channels, as is done in utterance-level separation, is not possible due to $\numspk > \numchn$ in general.
Instead, the placement of separated utterances on output channels is chosen such that no two utterances overlap in a channel.


\section{Variants of SDR}

A commonly used loss function for training of time-domain source separation networks is the \gls{SDR} \cite{Luo2018_TaSNetTimeDomainAudio,Luo2020_DualPathRNNEfficient,Heitkaemper2020_DemystifyingTasNetDissecting, LeRoux2019_SDRHalfbakedWell}.
It is generally defined in the logarithmic domain as
\def\SDR{\text{SDR}}
\begin{align}
    \loss{\SDR} (\estimate, \target) = -10 \log_{10} \frac{\norm{\target}^2}{\norm{\target - \estimate}^2}.
    \label{eq:sdr}
\end{align}

As a loss function for a source separator, output channels are usually treated independently. 
The loss is aggregated as the average over the output channels \cite{Luo2018_TaSNetTimeDomainAudio}. Neglecting the permutation ambiguity for the moment, this gives the averaged SDR (\aSDR)
\begin{align}
    \loss{\aSDR} (\estimatematrix, \targetmatrix) = -\frac{1}{\numchn} \sum_{\chnidx=1}^\numchn 10 \log_{10} \frac{\norm{\target_\chnidx}^2}{\norm{\target_\chnidx - \estimate_\chnidx}^2}.
    \label{eq:a-sdr}
\end{align}
%
Although $\loss{\aSDR}$ is commonly used, it becomes problematic in certain cases, e.g., for low volume or short utterances \cite{vonNeumann2021_GraphPITEnablingSegmentless}.
It does not allow speeding up Graph-PIT as we discuss later in \cref{sec:graph-pit}.
These problems can be eliminated by, instead of averaging the SDRs, summing the energies of the target and error signals \cite{vonNeumann2021_GraphPITEnablingSegmentless}:
\begin{align}
    \label{eq:sa-sdr}
    \loss{\saSDR} (\estimatematrix, \targetmatrix)& = -10\log_{10} \frac{\sum_{\chnidx=1}^\numchn\norm{\target_\chnidx}^2}{\sum_{\chnidx=1}^\numchn \norm{\target_\chnidx - \estimate_\chnidx}^2}. 
\end{align}
We call \cref{eq:sa-sdr} the source-aggregated SDR (\saSDR).
Similar modifications are possible for other objectives, e.g., SI-SDR \cite{LeRoux2019_SDRHalfbakedWell} or log-MSE \cite{Heitkaemper2020_DemystifyingTasNetDissecting}, but we restrict ourselves to the \gls{SDR} here.
We write $\sum_{\chnidx=1}^\numchn\norm{\target_\chnidx}^2$ as a matrix operation $\Tr(\targetmatrix\tran\targetmatrix)$ in the following equations.

\section{Speeding up utterance-level PIT}

One approach to utterance-level separation is \gls{uPIT} \cite{Kolbaek2017_MultitalkerSpeechSeparationa}, where the network outputs one signal for each speaker.
We assume for simplicity that $\numchn=\numspk=\numutt$ so that the signals $\target_\uttidx$ can be directly used as targets.
The permutation between speakers and output channels is ambiguous during training; \gls{uPIT} postulates to choose the permutation that yields the smallest loss.
It is often defined for losses $\L^\text{(pair)}$, such as \cref{eq:sdr}, that sum over losses between pairs of targets and estimates \cite{Kolbaek2017_MultitalkerSpeechSeparationa}
\begin{align}
    \jloss{uPIT}(\estimatematrix, \targetmatrix) = \min_{\perm\in\permsetf} \sum_{\chnidx=1}^\numchn \L^\text{(pair)}(\estimate_\chnidx, \target_{\perm(\chnidx)}),
    \label{eq:upit-sum}
\end{align}
where $\permsetf$ is the set of all permutations of length $\numchn$.
This formulation is not compatible with $\loss{\saSDR}$ in \cref{eq:sa-sdr}.
A more general definition can be found with a permutation matrix \cite{Tichavsky2004_OptimalPairingSignal}:
\begin{align}
    \jloss{uPIT}(\estimatematrix, \targetmatrix) = \min_{\permmatrix\in\permset_\numchn} \L(\estimatematrix, \targetmatrix\permmatrix),
    \label{eq:upit}
\end{align}
where $\L$ is a loss function over multiple pairs of targets and estimations, such as \cref{eq:a-sdr,eq:sa-sdr}, and $\permset_\numchn$ is the set of all permutation matrices $\permmatrix \in \{0,1\}^{\numchn\times\numchn}$. 
All entries in $\permmatrix$ are $0$ except for exactly one $1$ in each row and each column.

The permutation matrix $\hat{\permmatrix}$ that minimizes \cref{eq:upit} can be found naively by computing the loss for all possible permutations and selecting the one that yields the smallest loss.
This has a complexity of $\mathcal{O}(\numchn!)$ because $|\permset_\numchn| = \numchn!$ \cite{Kolbaek2017_MultitalkerSpeechSeparationa}, so it is only applicable for relatively small numbers of speakers.


If, for a certain $\L$, \gls{uPIT} can be expressed with \cref{eq:upit}, we can find a score matrix $\lossmatrix\in\mathbb{R}^{\numchn\times\numchn}$ with the elements $\lossmatrixentry_{i,j}=\L^\text{(pair)}(\estimate_i, \target_j)$ so that \cite{Tichavsky2004_OptimalPairingSignal}
\begin{align}
    \jloss{uPIT}(\estimatematrix, \targetmatrix) 
    = \min_{\perm\in\permsetf} \sum_{\chnidx=1}^C \L^\text{(pair)}(\estimate_\chnidx,\target_{\perm(\chnidx)}) = \min_{\permmatrix\in\permset_\numchn} \Tr(\lossmatrix\permmatrix) .
    \label{eq:upit-decomp-pair}
\end{align}
Finding the permutation matrix $\hat{\permmatrix}$ that minimizes \cref{eq:upit-decomp-pair} becomes a linear sum assignment problem \cite{Tichavsky2004_OptimalPairingSignal}, i.e., it is equivalent to selecting exactly one entry in each row and column of $\lossmatrix$ such that their sum is minimized.
This can be solved by the Hungarian algorithm in $\bigo(\numchn^3)$ time \cite{Kuhn_HungarianMethodAssignment,Munkres1957_AlgorithmsAssignmentTransportation,Tichavsky2004_OptimalPairingSignal}.

If a representation as in \cref{eq:upit-decomp-pair} is not possible, e.g., as for \cref{eq:sa-sdr}, we can still find $\hat{\permmatrix}$ with the Hungarian algorithm if $\J$ can be expressed as\footnote{The decomposition \cref{eq:upit2} does in general not exist, but it exists for most relevant separation objectives we are aware of.}
\begin{align}
    \jloss{uPIT}(\estimatematrix, \targetmatrix) = \outerfn(\min_{\permmatrix\in\permset_\numchn} \Tr(\lossmatrix\permmatrix),\estimatematrix,\targetmatrix),
    \label{eq:upit2}
\end{align}
with $\lossmatrix\in\mathbb{R}^{\numchn\times\numchn}$ and a strictly monotonously increasing function $\outerfn$, as we will demonstrate in the following.
The idea for the $\outerfn$ follows argmax rules and the inner representation for the permutation problem is known from \cite{Tichavsky2004_OptimalPairingSignal}.
\cref{eq:upit2} equals \cref{eq:upit-decomp-pair} for $\outerfn(x,\estimatematrix,\targetmatrix)=x$ and $\lossmatrixentry_{i,j}=\L^\text{(pair)}(\estimate_i, \target_j)$ with ${x=\min_{\permmatrix\in\permset_\numchn}\Tr(\lossmatrix\permmatrix)}$.
The decomposition is often not unique and the choice of $\outerfn$ and $\lossmatrix$ impacts the computational cost.
Further, some decompositions that can be found for \gls{uPIT} are not usable for Graph-PIT.

\paragraph{Decomposing the Signal-to-Distortion Ratio (SDR)}

The averaged SDR, \cref{eq:a-sdr}, can be expressed with $\outerfn(x,\estimatematrix,\targetmatrix)=x$ and $\lossmatrixentry_{i,j}=\SDR(\estimate_i,\target_j)$. For the alternative,
 $\loss{\saSDR}$, we get
\begin{align}
  \jloss{uPIT} &= \min_{\permmatrix\in\permset_\numchn} -10\log_{10} \frac{\Tr(\targetmatrix\tran\targetmatrix)}{\Tr((\targetmatrix\permmatrix - \estimatematrix)\tran(\targetmatrix\permmatrix - \estimatematrix))} \\
  &= -10\log_{10} \frac{\Tr(\targetmatrix\tran\targetmatrix)}{\min\limits_{\permmatrix\in\permset_\numchn}\Tr((\targetmatrix\permmatrix - \estimatematrix)\tran(\targetmatrix\permmatrix - \estimatematrix))}. \label{eq:decomp-sdr-1}
\end{align}
Although \cref{eq:decomp-sdr-1} looks complicated due to the matrix notation, the denominator is the \gls{MSE} between targets and permuted estimations.
The permutation $\permmatrix$ can be found by minimizing the \gls{MSE}. 
Comparing with \cref{eq:upit2} and factorizing the \gls{MSE}, we get
\begin{align}
    \outerfn^\text{(\saSDR-MSE)}(x,\estimatematrix,\targetmatrix)&=-10\log_{10}\frac{\Tr(\targetmatrix\tran\targetmatrix)}{\time\cdot x}\nonumber\\
    &=-10\log_{10}\frac{\sum_{\chnidx=1}^\numchn\norm{\target_\chnidx}^2}{\time\cdot x}\label{eq:f-sdr3-1},\\
    \lossmatrixentry^\text{(\saSDR-MSE)}_{i,j} &= \frac{1}{\time}\norm{\target_i - \estimate_j}^2, \label{eq:l-sdr3-1}
\end{align}
%
where $\time$ is the time length of the signals.
We can further simplify the denominator in \cref{eq:decomp-sdr-1} to find
%
%
%
\begin{align}
    \outerfn^\text{(\saSDR-dot)}(x, \estimatematrix, \targetmatrix) &= -10\log_{10}\frac{\Tr(\targetmatrix\tran\targetmatrix)}{\Tr(\targetmatrix\tran\targetmatrix) + \Tr(\estimatematrix\tran\estimatematrix) + 2x},\label{eq:f-sdr3} \\
    \lossmatrix^\text{(\saSDR-dot)} &= -\estimatematrix\tran\targetmatrix. \label{eq:l-sdr3}
\end{align}
The elements of $\lossmatrix$ are the dot products between pairs of estimations and targets $\lossmatrixentry_{i,j}=\estimate_i\rawtran\target_j$.
They are computed with a matrix multiplication which is efficient on modern computer hardware.
We show in \cref{sec:graph-pit} that the decomposition in \cref{eq:f-sdr3,eq:l-sdr3} is required for speeding up Graph-PIT.
%
%
%



\section{Speeding up Graph-PIT}
\label{sec:graph-pit}

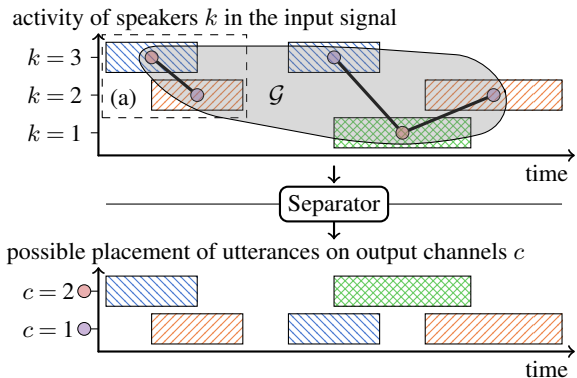
\begin{figure}[t]
    \centering
    \begin{tikzpicture}
    \def\t{0.2}
    \newcommand{\rect}[3]{($(c#2|-r#1) + (0, \t)$) rectangle ($(c#3|-r#1) + (0, -\t)$)}

    \path (0, 0) --coordinate[pos=0](c0) coordinate[pos=1/10](c1) coordinate[pos=2/10](c2) coordinate[pos=3/10](c3) coordinate[pos=4/10](c4) coordinate[pos=5/10](c5) coordinate[pos=6/10](c6) coordinate[pos=7/10](c7) coordinate[pos=8/10](c8) coordinate[pos=1](c9) (6, 0);

    \path (0, 0) --coordinate[pos=0](r0) coordinate[pos=1/2](r1) coordinate[pos=1](r2) (0, -1);
    \node at ($(r0-|c0) + (1.3,0.5)$) {activity of speakers $\spkidx$ in the input signal};
    \draw[black] ($(r0-|c0) + (-0.1,0)$) -- ++(-0.1, 0) node[left]{$\spkidx=3$};
    \draw[black] ($(r1-|c0) + (-0.1,0)$) -- ++(-0.1, 0) node[left]{$\spkidx=2$};
    \draw[black] ($(r2-|c0) + (-0.1,0)$) -- ++(-0.1, 0) node[left]{$\spkidx=1$};

    \draw[pattern1] \rect{0}{0}{2} node[pos=0.5,vertex,fill=palette-4](v11){};
    \draw[pattern1] \rect{0}{4}{6} node[pos=0.5,vertex,fill=palette-5](v12){};

    \draw[pattern2] \rect{1}{1}{3} node[pos=0.5,vertex,fill=palette-5](v21){};
    \draw[pattern2] \rect{1}{7}{9} node[pos=0.5,vertex,fill=palette-5](v22){};
    
    \draw[pattern3] \rect{2}{5}{8} node[pos=0.5,vertex,fill=palette-4](v31){};

    \node[draw,dashed,fit={(r0-|c0)(r1-|c3)},inner ysep=3mm, inner xsep=0.5mm](box){};
    \node[anchor=south west,inner sep=1pt,xshift=2pt,yshift=2pt] at (box.south west) {(a)};

    \draw[edge] (v11) -- (v21);
    \draw[edge] (v12) -- (v31) -- (v22);

    \draw[fill=gray, fill opacity=0.3, rounded corners=0.5cm] ($(v11) + (-0.4,0.15)$) -- ($(v12) + (0, 0.15)$) -- ($(v22) + (0, 0.5)$) --  ($(v22) + (0.3, -0.4)$) -- ($(v31) + (0.5, -0.15)$) -- ($(v31) + (-0.5, -0.15)$) -- ($(v21) + (-0.2, -0.2)$) -- cycle;
    \path (v11) -- (v31) node [pos=0.5,text=black]{$\graph$};

    \path (r2) -- ++(0, -2.1) -- coordinate[pos=0] (r3) coordinate[pos=1] (r4) ++(0, -0.5);
    \node at ($(r3-|c0) + (2.1,0.5)$) {possible placement of utterances on output channels $\chnidx$};
    \draw[black] ($(r3-|c0) + (-0.1,0)$) -- ++(-0.1, 0) node[left,xshift=-1mm]{$\chnidx=2$}node[left,vertex,fill=palette-4]{};
    \draw[black] ($(r4-|c0) + (-0.1,0)$) -- ++(-0.1, 0) node[left,xshift=-1mm]{$\chnidx=1$}node[left,vertex,fill=palette-5]{};

    \draw[pattern1] \rect{3}{0}{2};
    \draw[pattern1] \rect{4}{4}{6};
    \draw[pattern2] \rect{4}{1}{3};
    \draw[pattern2] \rect{4}{7}{9};
    \draw[pattern3] \rect{3}{5}{8};
    
    \draw ($(r2-|c0)!.45!(r3-|c0)$) -- ($(r2-|c9)!.45!(r3-|c9)$) node[box,pos=0.5,minimum height=0](sep){Separator};
    \draw[arrow] ($(sep) + (0,0.5)$) -- (sep);
    \draw[arrow] (sep) -- ($(sep) - (0,0.5)$);
    
    \draw[arrow] ($(r2-|c0) + (-0.1,-0.3)$) -- ($(r2-|c9) + (0.2,-0.3)$)  node[pos=1, below, anchor=north east]{time};
    \draw[arrow] ($(r2-|c0) + (-0.1,-0.3)$) -- ($(r0-|c0) + (-0.1,0.35)$);
    
    \draw[arrow] ($(r4-|c0) + (-0.1,-0.3)$) -- ($(r4-|c9) + (0.2,-0.3)$)  node[pos=1, below, anchor=north east]{time};
    \draw[arrow] ($(r4-|c0) + (-0.1,-0.3)$) -- ($(r3-|c0) + (-0.1,0.35)$);

\end{tikzpicture}
    \caption{Example of processing a three-speaker scenario using Graph-PIT with a two-output separator.
        Each box represents one utterance.
        \emph{Top}: Utterances in the meeting and the colored overlap graph $\graph$.  Graph-PIT is equivalent to \gls{uPIT} for an activity pattern as marked with (a).
        \emph{Bottom}: A possible assignment of utterances to output channels. Taken from \cite{vonNeumann2021_GraphPITGeneralizedPermutation}.}
    \label{fig:gpit}
\end{figure}

Graph-PIT \cite{vonNeumann2021_GraphPITGeneralizedPermutation} solves the meeting-level separation problem under the assumption that never more than $\numchn$ speakers are active at the same time, but in general $\numchn < \numspk$.
It allows placement of utterances from different speakers on the same output channel and distributing utterances from the same speaker across different output channels as long as they never overlap during training.

Using the matrix notation, we can define Graph-PIT as a natural extension of \gls{uPIT} (\cref{eq:upit}):
\begin{align}
    \jloss{Graph-PIT} = \min_{\coloringmatrix\in\coloringset} \L(\estimatematrix,\targetmatrix\coloringmatrix).
    \label{eq:gpit}
\end{align}
The matrix $\permmatrix\in\{0,1\}^{\numutt\times\numchn}$ is no longer a square permutation matrix but an assignment matrix that assigns utterances to output channels, i.e., several utterances can be assigned to each output channel.
This means that each row in $\permmatrix$ contains exactly one $1$.
The term $\targetmatrix\coloringmatrix\in\mathbb{R}^{\time\times\numchn}$ represents the target signals when the utterances are assigned to output channels with $\coloringmatrix$.
The columns of $\targetmatrix\coloringmatrix$ are no longer only permutations of $\targetmatrix$ but disjoint sums of utterance signals from $\targetmatrix$.

We have to find the set $\coloringset$ of matrices that are valid assignments of utterances to output channels, i.e., so that no two utterances overlap on the same output channel.
This is equivalent to a graph vertex coloring problem of the undirected graph $\graph=(\vertexset, \edgeset)$, whose vertices $\vertexset$ represent utterances and edges $\edgeset$ overlaps between two utterances.
Such a coloring is shown for an example graph in \cref{fig:gpit}.
The assignment of a color to a vertex represents the assignment of an utterance to an output channel.
An edge is added between two vertices if the corresponding utterances overlap in time:
\begin{align}
    \vertexset &= \{1,...,\numutt\},\\
    \edgeset &= \{\{u,v\}: u,v\in\vertexset, \text{utterances } u \text{ and } v \text{ overlap}\}.
\end{align}
%
Its adjacency matrix $\adjacencymatrix\in\{0,1\}^{\numutt\times\numutt}$ is given by its elements
\begin{align}
    a_{u,v} = \begin{cases}
        1,& \text{if } \{u,v\} \in \edgeset,\\
        0,& \text{otherwise}.
    \end{cases}
\end{align}

A vertex coloring is usually represented as a function $\pi:\vertexset \rightarrow \{1,...,\numchn\}$ so that $\pi(u)\neq\pi(v)$ if $\{u,v\}\in\edgeset$ \cite{Bollobas1979_Colouring}.
To be consistent with \cref{eq:upit}, we represent a coloring with the assignment matrix $\coloringmatrix\in\{0,1\}^{\numutt\times\numchn}$, where its entries are $p_{i,\pi(i)}=1$ for $i \in \{1,...,\numutt\}$ and all other entries are $0$.
If $\coloringmatrix$ is a valid coloring it satisfies
\begin{align}
    \Tr(\coloringmatrix\tran\adjacencymatrix\coloringmatrix) = 0.
\end{align}
We denote by $\coloringset$ the set of all valid $\numchn$-colorings of $\graph$.

%

Similar to \cref{eq:upit2} , we are interested in a decomposition for Graph-PIT where $\L$ can be expressed as a computation of a score matrix $\lossmatrix\in\mathbb{R}^{\numchn\times\numutt}$ and 
a strictly monotonously increasing $\outerfn$:
\begin{align}
    \jloss{Graph-PIT}(\estimatematrix, \targetmatrix) = \outerfn(\min_{\permmatrix\in\coloringset} \Tr(\lossmatrix\permmatrix),\estimatematrix,\targetmatrix).
    \label{eq:gpit-lossmatrix}
\end{align}

Note that $\coloringmatrix$ is no longer square and that some or all target signals $\targetmatrix\coloringmatrix$ in \cref{eq:gpit} are now the sum of multiple utterances.
Because of this we have to carefully select our objective since we no longer know a decomposition for every relevant source separation objective, as it was the case for \gls{uPIT}.
We specifically introduced $\loss\saSDR$ because we are not aware of a decomposition for $\loss\aSDR$.
Further, while \cref{eq:f-sdr3-1,eq:l-sdr3-1} work for \gls{uPIT}, this decomposition is not applicable for Graph-PIT because the \gls{MSE} cannot be factorized if the targets are the sums of multiple target utterances.
The matrix product in \cref{eq:f-sdr3,eq:l-sdr3}, however, works, as it directly uses utterance signals $\targetmatrix$ instead of the target sum signals $\targetmatrix\coloringmatrix$.

\subsection{Finding the assignment matrix}
\label{sec:assignment-algorithms}

Finding $\hat{\coloringmatrix}\in\coloringset$ that minimizes \cref{eq:gpit} for a general loss function $\L$ requires computing the full loss for all $\coloringmatrix\in\coloringset$.
The decomposition in \cref{eq:gpit-lossmatrix} on the other hand, if it exists, allows for more efficient search, where the assignment is solved on $\lossmatrix$ instead of computing the objective for each assignment.
If $\lossmatrix$ is treated as a score matrix assigning a score to each pair of vertex and color, $\hat{\coloringmatrix}$ is the coloring of $\graph$ with the minimal score.
Compared to \gls{uPIT}, the problem of finding $\hat{\coloringmatrix}$ is no longer a balanced linear sum assignment problem, since multiple utterances can be assigned to a single output channel.
The overlap graph further imposes constraints.
This prevents us from using the Hungarian algorithm.
The following sub-sections discuss different approaches for finding $\hat{\coloringmatrix}$.

\subsubsection{Optimal: Brute-force exhaustive search}

The naive way for solving the assignment problem is brute-force enumeration of all possible solutions and selecting the best one.
This assignment algorithm is used in \cite{vonNeumann2021_GraphPITGeneralizedPermutation}.
It has an exponential complexity limited by the number of colorings  $\bigo(\numchn(\numchn-1)^{\numutt-1})$.




\subsubsection{Greedy: Depth First Search}

A greedy but not necessarily optimal solution can be found with (incomplete) \gls{DFS} \cite{Even1979_GraphAlgorithms} in the space of partial solutions.
The algorithm colors vertices sequentially so that each newly selected color adds the smallest possible score, respecting the constraints imposed by the overlap graph.
If at any point a vertex cannot be colored, the latest coloring is undone and the next  possible color adding the smallest score is tested.
This algorithm has a best-case complexity of $\bigo(\numutt\numchn)$.


\subsubsection{Optimal: Branch-and-Bound Search}

The solution space can be searched in a more elegant way by using a branch-and-bound \cite{Land2010_AutomaticMethodSolving} algorithm to always find the optimal assignment.
The algorithm works in the same way as the \gls{DFS} algorithm, but continues search until the best solution is found.
During the search process, partial solutions that have a worse score than the currently best solution are discarded immediately as extending those can never yield the lowest score.

\subsubsection{Optimal: Dynamic Programming}
\label{sec:dynamic-programming}

\def\vgraph{\tilde{\graph}}
\def\vvertexset{\tilde{\vertexset}}
\def\vedgeset{\tilde{\edgeset}}
\def\uttset{\mathcal{U}_\uttidx}
\def\vuttset{\mathcal{U}^\text{(vis)}_\uttidx}
\def\Nu{\ensuremath{\mathcal{N}_u}}
\def\Nubar{\ensuremath{\bar{\mathcal{N}}_u}}
\def\C{\ensuremath{Q}}

We can significantly reduce the number of considered partial colorings by the use of \gls{DP} \cite{Dreyfus2002_RichardBellmanBirth}.
Given our scenario, we can sort the utterances $\uttidx$ by their start times and traverse $\graph$ in that order from the first utterance $\uttidx=1$ to the last one $\uttidx=\numutt$.
Let us denote the set of already visited utterances $v<u$ that overlap with $\uttidx$ as $\mathcal{N}_u=\{v: v<u, \{u,v\}\in\edgeset\}$.
Generally, we have to consider all possible valid colorings of $\{v\in\vertexset: v<u\}$ in the $\uttidx$-th step.
It turns out to be enough to only remember one coloring, the one with the best accumulated score, for each set of colorings that share the same colors of the utterances in \Nu, due to the structure of $\graph$.
Let us call this set of colorings obtained in the $u$-th step $\C_{u-1}$.
We construct an intermediate set $\tilde{\C}_u$ by extending each coloring in $\C_{u-1}$ with each valid color $c$ of $u$.
The colors $c$ must be unequal to all colors of the utterances in \Nu for this coloring because $u$ overlaps with all utterances in \Nu.
The accumulated score is increased by $\lossmatrixentry_{c,u}$.
The set $\C_u$, for the next step, is obtained from $\tilde{\C}_u$ by only keeping the coloring with the best accumulated score for each set of colorings that share the same coloring of $\mathcal{N}_{u+1}$.
The colorings are extended by one vertex in every step, so that the best coloring of $\graph$ is obtained after $U$ steps as the best coloring in $\C_U$.

The algorithm always performs $U$ steps and $|\C_u|\leq|\tilde{\C}_u| \leq C^{C-1}$ because we consider colorings with $C$ colors of $|\mathcal{N}_u| < C$ vertices in $\C$. 
We know that $|\mathcal{N}_u| < C$ because we assume that at most $C$ utterances overlap at a time.
The complexity $\bigo(UC^{C-1})$ is thus overall linear in the number of utterances.

\subsection{General remarks on speeding up}

A straightforward optimization method for graph coloring is to color each connected component of $\graph$ independently.
Since there cannot be an edge between two vertices in different connected components, the assignment for one of them does not influence the others.
The number of utterances in one connected component is typically limited, so that the complexity becomes linear in the number of connected components. It stays unaffected within a connected component, e.g., exponential.
This acceleration can be combined with any of the above algorithms.
Note that the \gls{DP} assignment algorithm from \cref{sec:dynamic-programming} does this implicitly.

\section{Experiments}


\label{sec:model}

We use a \gls{DPRNN}-TasNet \cite{Luo2020_DualPathRNNEfficient} as the separator.
Its configuration is kept similar to \cite{Luo2020_DualPathRNNEfficient} except for the depth.
Our network uses three stacked blocks instead of six, to keep the computational cost for training low \cite{vonNeumann2021_GraphPITGeneralizedPermutation}.

\subsection{Separation performance}


To validate that $\loss{\saSDR}$ does not  impair the performance compared to $\loss{\aSDR}$, we compare their performance in terms of SDR \cite{Fevotte2005_BSSEVALToolbox} for fully overlapped data from the WSJ0-2mix database \cite{Hershey2016_DeepClusteringDiscriminative} and \gls{WER} for artificially generated meeting-like data.
We use the same artificially generated meetings based on WSJ \cite{Garofolo1993_CsriWsj0Complete} as \cite{vonNeumann2021_GraphPITGeneralizedPermutation}.
Each meeting is roughly \SI{120}{\second} long and comprises 5 to 8 speakers, so that the overall speaker distribution is uniform across all meetings.
The meetings have an overlap ratio of \SIrange{0.2}{0.4}{} and utterances are selected uniformly.
A logarithmic weight between \SIrange{0}{5}{\decibel} is applied to all utterances to simulate volume differences.
We use a sample rate of \SI{8}{\kilo\hertz}.



We observed a slight improvement in \gls{SDR} for WSJ0-2mix from \SI{15.7}{\decibel} to \SI{15.9}{\decibel} when switching from $\loss{\aSDR}$ to $\loss{\saSDR}$.
The \gls{WER} improved for the meeting-like data from \SI{13.7}{\percent} to \SI{13.4}{\percent}.
This confirms that the proposed modification of the loss does not harm the performance.



\subsection{Run-time evaluation  
}

The run-time evaluation is performed on a GTX1080 graphics card, while the assignment algorithms are executed on the CPU.
Graph-PIT algorithms are implemented in Python.
The run-times shown here are averaged over 500 measurements.

\subsubsection{uPIT}

\begin{figure}[t]
    \centering
\begin{tikzpicture}

\begin{axis}[
legend cell align={left},
legend columns=2,
legend style={
  at={(0.5,1.05)},
  anchor=south,
},
log basis y={10},
xlabel={Number of speakers},
xmin=2, xmax=100,
ylabel={Runtime in s},
ymin=0.000257916411632708, ymax=1,
ymode=log,
xmode=log,
xtick={2,3,4,5,6,7,8,9,10,25,50,100},
xticklabels={2,3,4,5,6,7,8,,10,25,50,100},
grid=both,
height=4.7cm,
width=8cm,
grid style={line width=.2pt,draw=black!20},
tick style={color=black!30},
]
\addplot [very thick, palette-1, mark=*]
table {%
2 0.00111956313252449
3 0.00372272033244371
4 0.0117215533554554
5 0.0595480513945222
6 0.419197091460228
7 2.69138953387737
};
\addlegendentry{\cref{eq:upit-sum} Brute-force}
\addplot [very thick, palette-2, mark=o]
table {%
2 0.000663285851478577
3 0.00146951999515295
4 0.00309146750718355
5 0.0129929865896702
6 0.0930522194132209
7 0.573629681318998
8 4.61374649200588
};
\addlegendentry{\cref{eq:upit-decomp-pair} Brute-force}
\addplot [very thick, palette-3, mark=|, mark size=2pt]
table {%
2 0.000655239820480347
3 0.000808488205075264
4 0.000835263840854168
5 0.000946060456335545
6 0.00132812894880772
7 0.00113407962024212
8 0.00126714676618576
9 0.00132441911846399
10 0.00142092421650887
11 0.00152055278420448
12 0.00161129225045443
13 0.00171030290424824
14 0.00180688861757517
15 0.0019049009308219
16 0.0019960730150342
17 0.00210222192108631
18 0.00219299901276827
19 0.00229215707629919
20 0.0023791678622365
21 0.00248984880745411
22 0.0029662187024951
23 0.0030181410163641
24 0.00342586301267147
25 0.00293309591710567
26 0.00298894893378019
27 0.00309941247105598
28 0.00320117592811584
29 0.00371295668184757
30 0.00428972396999598
31 0.00441458951681852
32 0.00394682489335537
33 0.00466494981199503
34 0.00406773068010807
35 0.00392909850925207
36 0.00399026390165091
37 0.00410417769104242
38 0.00424038097262383
39 0.00444839790463448
40 0.00465643186122179
41 0.00487575523555279
42 0.00508070021867752
43 0.00528853673487902
44 0.00551757659763098
45 0.00573863431811333
46 0.00597101453691721
47 0.00622899122536182
48 0.00644759207963944
49 0.00671359684318304
50 0.00696662895381451
51 0.00722546681761742
52 0.00749153301119804
53 0.0077349766716361
54 0.00804418683052063
55 0.00829769138246775
56 0.00858545530587435
57 0.00886336125433445
58 0.00918652165681124
59 0.00945148456841707
60 0.00973667684942484
61 0.0100587994605303
62 0.0104003477841616
63 0.0107340601086617
64 0.0111742464452982
65 0.0114757654815912
66 0.0118112963810563
67 0.0120476775243878
68 0.0124819265678525
69 0.0130974391847849
70 0.0134062726795673
71 0.0138800394907594
72 0.014240932688117
73 0.0146823197230697
74 0.0150981340557337
75 0.0154321897029877
76 0.0157564750686288
77 0.0162201898917556
78 0.0165394769608974
79 0.0169989151880145
80 0.0173867690935731
81 0.0179113453626633
82 0.0182171382009983
83 0.0185719924420118
84 0.0189201910048723
85 0.0193608831241727
86 0.0198166075348854
87 0.020251544713974
88 0.0207520397752523
89 0.0211671687662601
90 0.0216415830329061
91 0.0222319429367781
92 0.0225877267494798
93 0.0231085207685828
94 0.0234507482498884
95 0.024071258790791
96 0.0241181664913893
97 0.0247844294458628
98 0.0255522686615586
99 0.0260198545455933
};
\addlegendentry{\cref{eq:l-sdr3-1} Hungarian}
\addplot [very thick, palette-4, mark=Mercedes star,
mark options={scale=1, line width=0.5}
]
table {%
2 0.000781397297978401
3 0.000785539038479328
4 0.000768032222986221
5 0.000792359076440334
6 0.000960892029106617
7 0.000785854756832123
8 0.000795659683644772
9 0.000798098780214787
10 0.000800776518881321
11 0.000806332379579544
12 0.000803693234920502
13 0.000806606970727444
14 0.000814002752304077
15 0.000814467184245586
16 0.000810946710407734
17 0.000905516818165779
18 0.000915412493050098
19 0.000874980688095093
20 0.000847542397677898
21 0.000836061909794807
22 0.00100878827273846
23 0.000848195590078831
24 0.00101570654660463
25 0.000861965157091618
26 0.000851208530366421
27 0.000861808396875858
28 0.000860686041414738
29 0.00104114502668381
30 0.00104066543281078
31 0.000868039280176163
32 0.00105018757283688
33 0.00157624267041683
34 0.00138580091297626
35 0.00139066692441702
36 0.00139822248369455
37 0.0014001227542758
38 0.0014017540588975
39 0.00141525518149137
40 0.00140592899173498
41 0.00141240380704403
42 0.00140831302851439
43 0.00141609102487564
44 0.00143330283463001
45 0.00143512781709433
46 0.0014326062053442
47 0.00143964845687151
48 0.00143239237368107
49 0.00144881319254637
50 0.0014611329510808
51 0.00146912343800068
52 0.00145710933953524
53 0.00144460514187813
54 0.00148945096880198
55 0.0015147278085351
56 0.00149603847414255
57 0.00147565443068743
58 0.00149601329118013
59 0.00152828801423311
60 0.00150137644261122
61 0.00151481814682484
62 0.00152563776820898
63 0.00151097480207682
64 0.00177660275250673
65 0.00251227173954248
66 0.00248009871691465
67 0.00238854251801968
68 0.00244643282145262
69 0.0024954679235816
70 0.00249784387648106
71 0.00248744614422321
72 0.00254146326333284
73 0.00253212403506041
74 0.0023204206302762
75 0.00255817484110594
76 0.00249979939311743
77 0.00260399334132671
78 0.00247185312211514
79 0.00262288011610508
80 0.00253256287425756
81 0.0026533205807209
82 0.00252284463495016
83 0.00264215763658285
84 0.0026005644723773
85 0.00240559354424477
86 0.00247252099215984
87 0.00244344267994165
88 0.00254757635295391
89 0.00249251898378134
90 0.00267801024019718
91 0.0026478336006403
92 0.00275321368128061
93 0.00275562092661858
94 0.00271634090691805
95 0.00256066616624594
96 0.00245821721851826
97 0.00342012889683247
98 0.00360529497265816
99 0.00367751751095057
};
\addlegendentry{\cref{eq:l-sdr3} Hungarian}
\addplot[red, dashed] 
table {%
2 0.131374
3 0.1318
4 0.133198
5 0.12892
6 0.131052
7 0.128424
8 0.128523
9 0.130014
10 0.131683
25 0.146483
50 0.174603
100 0.246117
};
\node[anchor=south west, red] at (2,0.13) {\footnotesize DPRNN time};
\end{axis}

\end{tikzpicture}
    \caption{Run-time over number of speakers for different variants for computing the SDR for signals of \SI{4}{\second} length. The run-time required for a forward and backward pass of a corresponding DPRNN network is shown as a dashed red line.}
    \label{fig:uPIT-runtime}
\end{figure}
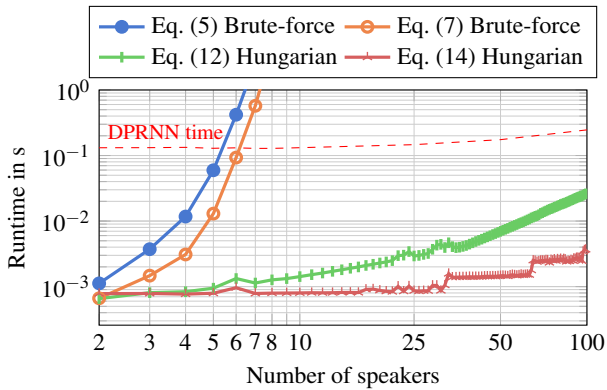

\cref{fig:uPIT-runtime} shows an analysis of the run-time of the different uPIT optimizations for $\loss{\saSDR}$.
The time required for a forward and backward step of a separation network as described in \cref{sec:model} is marked with a red dashed line.
Solving the permutation problem with a brute-force search has a factorial complexity and exceeds the run-time of the network already for $6$ speakers, even when most of the heavy-lifting is deferred to computation of a score matrix (\enquote{\cref{eq:upit-decomp-pair} Brute-force}).
Both variants that use the Hungarian algorithm are polynomial in time.
Their run-time is dominated by the computation of $\lossmatrix$.
Computing the \gls{MSE} for $\lossmatrix$ as in \cref{eq:l-sdr3-1} is slower than the dot product in \cref{eq:l-sdr3}. 
A carefully tuned low-level implementation could speed up the \gls{MSE} to be faster than the dot product because the squared error can be implemented without rather slow floating point multiplications. 
Since the run-times of both variants are more than one order of magnitude smaller than the time required for the network, such an implementation is not required.
The permutation problem can be solved for $100$ speakers in a fraction of a second with \cref{eq:l-sdr3} and the Hungarian algorithm, taking up a negligible amount of time in practical applications.


\subsubsection{Graph-PIT}

\begin{figure}[t]
    \centering
\begin{tikzpicture}

\definecolor{color0}{rgb}{0.282352941176471,0.470588235294118,0.815686274509804}
\definecolor{color1}{rgb}{0.933333333333333,0.52156862745098,0.290196078431373}
\definecolor{color2}{rgb}{0.415686274509804,0.8,0.392156862745098}
\definecolor{color3}{rgb}{0.83921568627451,0.372549019607843,0.372549019607843}
\definecolor{color4}{rgb}{0.584313725490196,0.423529411764706,0.705882352941177}

\begin{axis}[
legend cell align={left},
legend columns=2,
legend style={
  at={(0.5,1.05)},
  anchor=south,
},
log basis y={10},
xlabel={Number of utterances},
ylabel={Runtime in s},
ymode=log,
grid=both,
height=4.7cm,
width=8cm,
xmin=2, xmax=29,
ymin=3.9e-5, ymax=1,
max space between ticks=20,  
grid style={line width=.2pt,draw=black!20},
tick style={color=black!30},
]
\addplot [very thick, gray, dashed]
table {%
2 0.00116797767579556
3 0.00132768519222736
4 0.00122204475104809
5 0.00168215971440077
6 0.00185190834105015
7 0.0019374692440033
8 0.00196626894176006
9 0.00186342969536781
10 0.00252929851412773
11 0.00269922938197851
12 0.00287501286715269
13 0.00303982362151146
14 0.00308328554034233
15 0.00338926270604134
16 0.00356610048562288
17 0.00373788941651583
18 0.00381439264863729
19 0.00409234207123518
20 0.00426477808505297
21 0.00444218117743731
22 0.00462039776146412
23 0.00479734651744366
24 0.00497547313570976
25 0.0051560216024518
26 0.00533465027809143
27 0.00551351990550756
28 0.00569493111222982
29 0.0058649343624711
};
\addlegendentry{Computation of $\lossmatrix$}
\addplot [very thick, color1,mark=o]
table {%
2 0.00356427326798439
3 0.00785700265318155
4 0.01472092974931
5 0.0390237800404429
6 0.077560824342072
7 0.174468240477145
8 0.376247875913978
9 0.793966343328357
10 1.68352524802089
};
\addlegendentry{Unoptimized \cref{eq:gpit}}
\addplot [very thick, color4, mark = triangle]
table {%
2 0.000148715823888779
3 0.000165736898779869
4 0.000171216279268265
5 0.000265558771789074
6 0.000841398909687996
7 0.000844341404736042
8 0.00202753268182278
9 0.00398616507649422
10 0.00798677306622267
11 0.0164714854955673
12 0.0332208566740155
13 0.068723302334547
14 0.140531755313277
15 0.288550496287644
16 0.595131199546158
17 1.21552449833602
};
\addlegendentry{Brute-force}
\addplot [very thick, color3, mark=+]
table {%
2 0.000166797898709774
3 0.000228393711149693
4 0.000236125141382217
5 0.000303418152034283
6 0.000764160603284836
7 0.000470503084361553
8 0.000860301591455936
9 0.00123921114951372
10 0.00149316571652889
11 0.00210633233189583
12 0.00284488491714001
13 0.00361001331359148
14 0.00579340778291225
15 0.00734632458537817
16 0.010727463401854
17 0.0156545901298523
18 0.0209180035814643
19 0.0235766452178359
20 0.036761873960495
21 0.0487651356682181
22 0.0676024286821485
23 0.0663813169300556
24 0.116484025865793
25 0.135471763499081
26 0.310530985668302
27 0.265484242364764
28 0.278989633955061
29 0.90229992352426
};
\addlegendentry{Branch and bound}
\addplot [very thick, color0, mark=x]
table {%
2 0.000118579417467117
3 0.000139742977917194
4 0.000108639523386955
5 0.000135011970996857
6 0.000140545678138733
7 0.000159007757902145
8 0.000206614546477795
9 0.000220147147774696
10 0.000239440351724625
11 0.000252406820654869
12 0.00028076246380806
13 0.000289274416863919
14 0.000304659754037857
15 0.000321508124470711
16 0.00034041054546833
17 0.000357216373085976
18 0.000375628620386124
19 0.000388296619057655
20 0.000405502766370773
21 0.000543289817869663
22 0.000435959175229072
23 0.000530764199793339
24 0.000522950701415539
25 0.000317549407482147
26 0.000434096083045006
27 0.000516637712717056
28 0.000534276403486729
29 0.000554195307195186
};
\addlegendentry{DFS}
\addplot [very thick, color2,mark=diamond]
table {%
2 6.59798458218574e-05
3 9.51443985104561e-05
4 7.87652283906937e-05
5 9.7908116877079e-05
6 0.000103807635605335
7 0.000134680233895779
8 0.000179722271859646
9 0.000198966190218926
10 0.000220458544790745
11 0.000242447853088379
12 0.00026327058672905
13 0.000281030982732773
14 0.000310466550290584
15 0.00031452339142561
16 0.00033851183950901
17 0.000315884426236152
18 0.000375055968761444
19 0.000396823659539223
20 0.000416367724537849
21 0.000385524295270443
22 0.00046117328107357
23 0.000545360222458839
24 0.000496311374008655
25 0.000504512004554272
26 0.000806440226733685
27 0.000553402863442898
28 0.000578284561634064
29 0.000468288734555245
};
\addlegendentry{Dynamic programming}
\addplot [red, dashed]
table {%
2 0.195693
3 0.2382257
4 0.27892510000000004
5 0.3167707
6 0.3507175
7 0.3678388
8 0.43048279999999994
9 0.44137750000000003
10 0.49261520000000003
11 0.5350092
12 0.5626515000000001
13 0.6125605000000001
14 0.6389149
15 0.6694822
16 0.7111985
17 0.7289245
18 0.7580833
19 0.8070021
20 0.8953457
21 0.9437540999999999
22 0.9904498
23 1.0799341999999998
24 1.1210502999999998
25 1.2068229000000001
26 1.281529
27 1.3623123
28 1.4257358999999998
29 1.5335623
};
\node[anchor=south west, red] at (1.7,0.3) {\footnotesize DPRNN time};
\end{axis}

\end{tikzpicture}
    \caption{Comparison of the different assignment algorithms for different sizes of connected components on toy example data, computing \saSDR. All algorithms use a pre-computed score matrix (gray dashed line) except for the \enquote{Unoptimized} one. The separator has three outputs. All simulated utterances have a length of \SI{2}{\second} with \SI{0.5}{\second} overlap. The red dashed line indicates the time required for a forward and backward step of the separation network.}
    \label{fig:GPIT-runtime-num-utterances}
\end{figure}
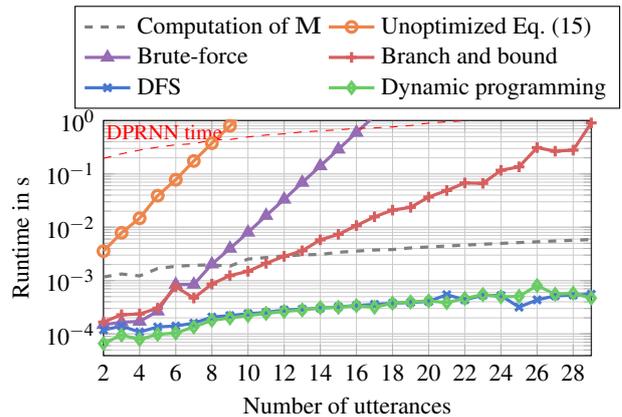

We compare the run-time of the different assignment algorithms for Graph-PIT in \cref{fig:GPIT-runtime-num-utterances} for a single connected component with varying numbers of utterances and a separator with three output channels.
The algorithms compute $\loss{\saSDR}$ since a decomposition for $\loss{\aSDR}$ is not possible for Graph-PIT.
All algorithms use a pre-computed loss matrix, except for the \enquote{Unoptimized} algorithm, which computes the full loss for all assignments (\cref{eq:gpit}).
The time required for computing $\lossmatrix$ is drawn with a dashed gray line.
Any algorithm that needs more time than the computation of $\lossmatrix$ dominates the overall run-time.

Both brute-force assignment algorithms (\enquote{Unoptimized} and \enquote{Brute-force}) show an exponential behavior and dominate the run-time for relatively small numbers of utterances.
The branch-and-bound algorithm has a better run-time, but is still exponential.
The run-times of the \gls{DP} and \gls{DFS} algorithms are linear in the number of utterances where \gls{DP} finds the optimal solution.
Their run-time is negligible compared to the computational cost required to compute $\lossmatrix$.
A connected component contains on average $3$ utterances in the CHiME-6 corpus \cite{Watanabe2020_CHiME6ChallengeTackling}, with a few outliers contain more than $30$ utterances.
These would dominate the run-time for all exponential-time algorithms; the \enquote{Unoptimized} algorithm, for example, takes more then \SI{40}{\second} for 15 utterances.

A further speedup could be realized by implementation in a lower-level language, such as C.
Our implementation of the \gls{DP} algorithm has a negligible run-time for practical examples.




\section{Conclusions}
We present a general framework for decomposing loss functions for efficiently finding the optimal assignment of targets to output channels in \gls{uPIT} or Graph-PIT training of a separation network.
We observe a large speedup compared to the naive implementations, so that the run-time of the assignment algorithm becomes negligible even for large numbers of speakers or utterances.

\section{Acknowledgements}
Computational resources were provided by the Paderborn Center for Parallel Computing.

\small
\bibliographystyle{ieeetr}
\balance
\bibliography{references}


\end{document}